\documentclass[modern]{aastex63}
\usepackage[utf8]{inputenc}
\usepackage{amsmath}

\newcommand{\sbu}{Department of Physics and Astronomy, Stony Brook University, Stony Brook NY 11794, United States}
\newcommand{\cca}{Center for Computational Astrophysics, Flatiron Institute, New York NY 10010, United States}

\newcommand{\ThankfulPulsar}{J0740+6620}
\newcommand{\mmedthankful}{2.03}
\newcommand{\mplusthankful}{0.10}
\newcommand{\mminusthankful}{0.08}

\begin{document}

\title{A Population-Informed Mass Estimate for Pulsar \ThankfulPulsar}
\author[0000-0003-1540-8562]{Will M. Farr}
\email{will.farr@stonybrook.edu}
\email{wfarr@flatironinstitute.org}
\affiliation{\sbu}
\affiliation{\cca}

\author{Katerina Chatziioannou}
\email{kchatziioannou@flatironinstitute.org}
\affiliation{\cca}

\begin{abstract}
Galactic double neutron star systems have a tight mass distribution around $\sim 1.35 M_{\odot}$, but the mass distribution of all
known pulsars is broader~\citep{Tauris:2017omb}. Here we reconstruct the \citet{Alsing2018,Antoniadis:2016hxz} bimodal mass distribution of pulsars observed in binary systems, incorporating data from observations of \ThankfulPulsar{} which were not available at the time of those works.  Because \ThankfulPulsar{} is an outlier in the mass distribution with non-negligible uncertainty in its mass measurement, its mass receives a large correction from the population, becoming $m_{\ThankfulPulsar{}} = \mmedthankful^{+\mplusthankful}_{-\mminusthankful} \, M_\odot$ (median and 68\% CI).  Stochastic samples from our population model, including population-informed pulsar mass estimates, are available at \url{https://github.com/farr/AlsingNSMassReplication}.
\end{abstract}

\section{The Pulsar Mass Distribution}

  There are three classes of pulsar mass measurements used in \citet{Alsing2018}.
The first consist of well-constrained pulsar mass measurements, where the likelihood function is assumed to be (proportional to) a Gaussian:
\begin{equation}
    p\left( d_p \mid m_p \right) \propto \exp\left( - \frac{\left( m_p - \mu \right)^2}{2 \sigma^2} \right),
\end{equation}
with $d_p$ the (abstract) pulsar measurement ``data,'' and $m_p$ the true mass of the pulsar.  Here $\mu$ is the reported mass measurement and $\sigma$ is its uncertainty.  The second class consists of measurements of the mass function,
\begin{equation}
    f \equiv \frac{m_p q^3 \sin^2 \iota}{\left( 1 + q\right)^2},
\end{equation}
where $q = m_c / m_p$ ($m_c$ is the companion mass; this implies $0 < q < \infty$) and $\iota$ is the angle between the orbital angular momentum and the line of sight; $f$ is assumed to be measured perfectly (uncertainties on $f$ are below the part-per-thousand level, and therefore ignorable); and measurements of the total mass of the binary system, $m_t = m_p \left( 1 + q \right)$ with an assumed Gaussian likelihood centered at the measured total mass, $\mu_t$, with standard deviation $\sigma_t$.  The complete likelihood is 
\begin{equation}
    p\left( d_p \mid m_p, m_t, \iota \right) \propto \delta\left( f\left( m_p, m_t, \iota \right) - f_p \right) \exp\left( -\frac{\left( m_t - \mu_t\right)^2}{2 \sigma_t^2}\right),
\end{equation}
with $f_p$ the measured mass function.  Integrating over $\iota$ with an isotropic prior (flat in $\cos \iota$) leaves the marginal likelihood 
\begin{equation}
    \label{eq:marginal-like-mt}
    p\left( d_p \mid m_p, m_t \right) \propto \exp\left( - \frac{\left( m_t - \mu_t \right)^2}{2 \sigma_t^2} \right) \frac{m_t^{4/3}}{3 \left( m_t - m_p \right)^2 f_p^{1/3} \sqrt{1 - \frac{f_p^{2/3} m_t^{4/3}}{\left(m_t - m_p\right)^2}}}
\end{equation}
(compare \citet{Alsing2018} Eq. (3)).  The third class consists of systems with (perfect) measurements of the mass function and Gaussian uncertainties on the mass ratio, $q$; similarly, integrating over $\iota$ with an isotropic prior leads to a marginal likelihood 
\begin{equation}
    p\left( d_p \mid m_p, q \right) \propto  \exp\left( - \frac{\left( q - \mu_q \right)^2}{2 \sigma_q^2} \right) \frac{\left( 1 + q \right)^{4/3}}{3 f_p^{1/3} m_p^{2/3} q^2 \sqrt{1 - \left(\frac{f_p}{m_p} \right)^{2/3} \frac{\left(1 + q\right)^{4/3}}{q^2}}};
\end{equation}
this is Eq.\ \eqref{eq:marginal-like-mt} with $m_t \to m_p \left( 1 + q \right)$ in the term that arises from integrating over $\iota$; notably, this \emph{differs} from \citet{Alsing2018} Eq.\ (4).

For our analysis, we extend the \citet{Alsing2018} data set to include the \citet{Cromartie2020} measurement of \ThankfulPulsar{} with $\mu = 2.14 \, M_\odot$ and $\sigma = 0.1 \, M_\odot$.  

We fit a pulsar population model to this data set with 
\begin{multline}
    \label{eq:pop-model}
    p\left( m_p \mid A, m_\mathrm{max}, \mu_{1,2}, \sigma_{1,2} \right) = \\ \begin{cases} 
        A \alpha \exp\left( - \frac{\left( m_p - \mu_1 \right)^2}{2 \sigma_1^2}\right) + \left(1 - A \right) \beta \exp\left( - \frac{\left( m_p - \mu_2 \right)^2}{2 \sigma_2^2} \right) & 0 < m_p \leq m_\mathrm{max} \\
        0 & \textnormal{otherwise}
    \end{cases},
\end{multline}
where $\alpha = \alpha\left( m_\mathrm{max}, \mu_1, \sigma_1 \right)$ and $\beta = \beta\left( m_\mathrm{max}, \mu_2, \sigma_2 \right)$ normalize the Gaussian terms in the sum to individually integrate to 1 for $0 < m_p \leq m_\mathrm{max}$ so that $A$ is the fraction of the population in the first Gaussian component.  This is equivalent to the \citet{Alsing2018} model with $n=2$ Gaussian components.  We impose that $\mu_1 < \mu_2$ to break labeling degeneracy.  We use the \texttt{Stan} sampler \citep{Carpenter2017} to sample over the pulsar population parameters, pulsar masses, and total masses or mass ratios (for the appropriate subset of pulsars). We use the same priors as \citet{Alsing2018} on the pulsar-level parameters.  Our code and samples, as well as the \LaTeX{} source for this document can be found at \url{https://github.com/farr/AlsingNSMassReplication}.

An example of our results for the pulsar population are found in the left panel of Fig. \ref{fig:pdfs}, where we show the distribution of pulsar masses marginalized over the posterior for the distribution parameters and the posterior on the $m_\mathrm{max}$ parameter; similarly to \citet{Alsing2018}, we find that the pulsar mass distribution tapers off for $m_p \gtrsim 2 \, M_\odot$, either because there is a sharp cutoff ($m_\mathrm{max} \simeq 2 \, M_\odot$), leading to a peak in the posterior for $m_\mathrm{max}$ near $2 \, M_\odot$, or because the second Gaussian component is narrow and $m_\mathrm{max}$ is unconstrained, leading to the ``tail'' in $m_\mathrm{max}$ running up to $3 \, M_\odot$.  We find somewhat weaker evidence for a sharp cutoff than reported in \citet{Alsing2018} with the Bayes Factor in favor of a cutoff varying between 1:1 and 5:1 depending on the choice of $m_\mathrm{max}$ prior.

\section{Updated Mass Estimate for J0740+6620}

\begin{figure}
    \plottwo{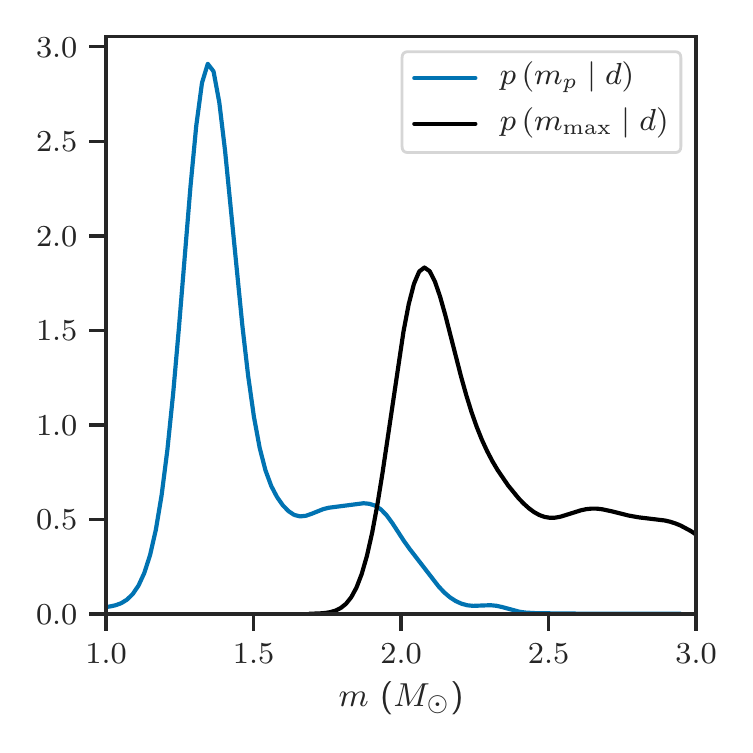}{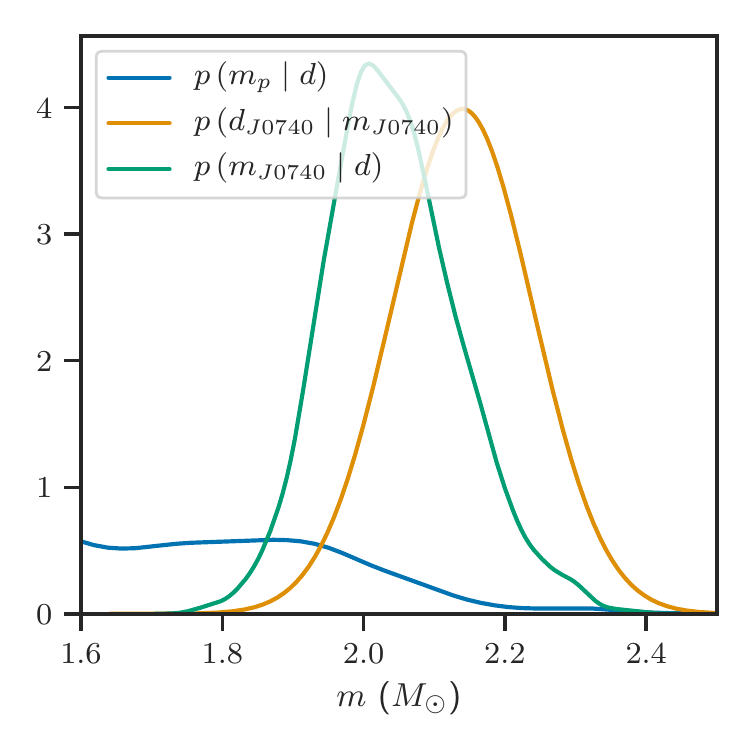}
    \caption{Left: Posterior predictive distribution---the mass distribution obtained by marginalizing over all pulsar parameters and population-level parameters in Eq.\ \eqref{eq:pop-model} given the complete data set---for the pulsar mass distribution (blue) and posterior for the maximum mass (black). Right: Likelihood for the mass of \ThankfulPulsar{} assuming only data for that pulsar (orange) and the posterior on its mass obtained from marginalization of the joint posterior on all pulsar parameters and population parameters from the population inference based on the entire data set of galactic pulsar mass measurements (green).  The tail of the pulsar mass distribution from the left figure is also shown in blue; because the tail tapers strongly over the range of masses supported by the \ThankfulPulsar{} likelihood the population-informed mass estimate is ``pulled'' to lower values.}
    \label{fig:pdfs}
\end{figure}

We report an updated mass estimate for \ThankfulPulsar{} informed by this population analysis.  The \citet{Cromartie2020} mass estimate is an outlier relative to the distribution, with an uncertainty $\sigma_m \simeq 0.1 \, M_\odot$ comparable to the overall pulsar distribution width $\sigma_2 \simeq 0.25 \, M_\odot$, and therefore receives a large ``correction'' from the joint distribution \citep{Fishbach2020}.  We find $m_{\ThankfulPulsar{}} = \mmedthankful{}^{+\mplusthankful{}}_{-\mminusthankful{}} \, M_\odot$ (median and symmetric 68\% credible interval) when incorporating a population model for galactic pulsars similar to \citet{Alsing2018}, right panel of Fig.~\ref{fig:pdfs}. The updated mass estimate for \ThankfulPulsar{} is informed by the population analysis of galactic pulsar mass measurements.  These include double neutron star binaries, neutron star-white dwarf binaries, and
X-ray binaries. Most heavy pulsars with precise mass measurements are in neutron star-white dwarf binaries; it is these systems that control the mass distribution near $m_p \simeq 2 \, M_\odot$.  \ThankfulPulsar{} is also of this type, so the systems with the strongest effect on the ``correction" to the \ThankfulPulsar{} mass are likely the most similar among the set of pulsars considered here. 

\software{{\tt matplotlib}~\citep{Hunter:2007}, {\tt stan}~\citep{JSSv076i01}, 
{\tt numpy}~\citep{numpy}, {\tt scipy}~\citep{2020SciPy-NMeth}, {\tt astropy}~\citep{astropy:2013, astropy:2018}, \texttt{arviz}~\citep{arviz_2019}, \texttt{pandas}~\citep{reback2020pandas}, \texttt{seaborn}~\citep{michael_waskom_2020_3767070}}

\acknowledgments
We thank Cole Miller for useful discussion.  WMF and KC are supported in part by the Simons Foundation.

\clearpage{}

\bibliography{ar,alsing_references}

\end{document}